\documentclass{article}
\makeatletter
\providecommand{\headeright}{}

\makeatother

\usepackage{arxiv}

\usepackage{parskip}
\usepackage{subcaption}
\usepackage{graphicx}
\usepackage{geometry}
\usepackage{hyperref}
\usepackage{float}

\usepackage{amsmath}
\usepackage{amsfonts}
\usepackage{amssymb}
\usepackage{bm}

\usepackage[utf8]{inputenc}
\usepackage[T1]{fontenc}
\usepackage{hyperref}
\usepackage{url}
\usepackage{booktabs}
\usepackage{amsfonts}
\usepackage{nicefrac}
\usepackage{microtype}
\usepackage{graphicx}
\usepackage[numbers]{natbib}
\usepackage{doi}
\usepackage{parskip}
\usepackage{changepage}
\usepackage[most]{tcolorbox}

\usepackage{subfiles}

\title{Capturing the Topological Phase Transition and Thermodynamics of the 2D XY Model via Manifold-Aware Score-Based Generative Modeling}

\author{{Pratyush Jha}\\
	Bengaluru, India\\
	\texttt{pratyushjha254@gmail.com} \\
}
\renewcommand{\headeright}{}

\begin{document}
\maketitle
\begin{abstract}
The application of generative modeling to many-body physics offers a promising pathway for analyzing high-dimensional state spaces of spin systems. However, unlike computer vision tasks where visual fidelity suffices, physical systems require the rigorous reproduction of higher-order statistical moments and thermodynamic quantities. While Score-Based Generative Models (SGMs) have emerged as a powerful tool, their standard formulation on Euclidean embedding space is ill-suited for continuous spin systems, where variables inherently reside on a manifold. In this work, we demonstrate that training on the Euclidean space compromises the model's ability to learn the target distribution as it prioritizes to learn the manifold constraints. We address this limitation by proposing the use of Manifold-Aware Score-Based Generative Modeling framework applied to the $64 \times 64$ 2D XY model (a 4096-dimensional torus). We show that our method estimates the theoretical Boltzmann score with superior precision compared to standard diffusion models. Consequently, we successfully capture the Berezinskii-Kosterlitz-Thouless (BKT) phase transition and accurately reproduce second-moment quantities, such as heat capacity without explicit feature engineering. Furthermore, we demonstrate zero-shot generalization to unseen lattice sizes, accurately recovering the physics of variable system scales without retraining. Since this approach bypasses domain-specific feature engineering, it remains intrinsically generalizable to other continuous spin systems.
\end{abstract}


\section{Introduction} 
The tractability of representing probability distributions in high-dimensional spaces is a fundamental challenge across scientific disciplines. In computer vision, a single datapoint is an image that comprises thousands of pixel variables, making the direct estimation of the underlying distribution computationally prohibitive. Similarly, in condensed matter physics, the size of the state space grows exponentially with the increase in the number of particles. \cite{MLPhasesOfMatter8} This exponential complexity precludes the use of conventional computational methods, necessitating the development of novel generative modeling techniques to analyze such many-body systems that can solve the dimensionality problem by implicitly learning the complex distribution and providing methods to sample from the learned distributions.

The application of generative models to physical systems requires the need to accurately estimate the rigorous constraints of those systems, which are otherwise less relevant in standard image synthesis tasks. While perceptual metrics effectively validate visual quality in computer vision tasks, modeling physical systems requires the precise reproduction of statistical moments. A generative model for spin systems must accurately capture not only the mean energy (first moment) but also quantities relying on higher-order moments such as the specific heat capacity. This requirement stems from the need for high explainability, where the model must satisfy the rigorous thermodynamic quantities of the underlying Hamiltonian rather than merely producing visually plausible spin structures.

Inspired by the success of deep learning in high-dimensional datasets, researchers have extensively explored its application to statistical physics. Early efforts focused on training models on classification tasks \cite{MLPhasesOfMatter8,classification1_21}. Ref. \cite{MLPhasesOfMatter8} demonstrated that convolutional neural networks could classify phases to identify phase transitions in the Ising Model and further performed rigorous analysis of this approach to modeling XY Model data \cite{MelkoMLvortices4}. Ref. \cite{GANContessi1} utilized entanglement spectra and Generative Adversarial Networks (GANs) \cite{GAN19} to detect the Berezinskii-Kosterlitz-Thouless (BKT) transition without prior knowledge of the order parameters using classification techniques. Building on these classification successes, the field has pivoted toward full generative modeling to replicate the system's statistics. Ref. \cite{SinghcGAN10} utilized conditional GANs to generate samples across different temperatures. Despite these advancements, faithfully reproducing the thermodynamic properties of the 2D XY model solely with a trained model remains a complex objective.

Multiple works have shown promising results by using data-specific architectural modifications that defeat the purpose of developing a generalized approach. It is essential to develop a generalized approach that eliminates the need for system-specific feature engineering \cite{MelkoMLvortices4, GANContessi1}. Such a strategy ensures that the resulting techniques are not merely tailored solutions for isolated cases, but rather mathematically rigorous and fundamentally robust methodologies applicable to a diverse class of many-body systems. Furthermore, methods that attempt to correct these discrepancies by explicitly modifying the loss function to match second-moment quantities often compromise on this generalizability. 

Recently, Score-Based Generative Models (SGMs) have superseded GANs and VAEs in many domains, driven by their superior training stability and robust mathematical foundations \cite{DiffusionBeatsGAN15}. In statistical physics, this progress has inspired applications to spin systems \cite{IsingDiffusion3,Diffusion1ForLattices14,vega2025group_18}. Ref. \cite{IsingDiffusion3} presents a comparative study between Denoising Diffusion Probabilistic Model (DDPM) and Generative Adversarial Networks (GAN) for the Ising system. Another approach is the use of Group-Equivariant Score Networks introduced in Ref. \cite{vega2025group_18}. Ref. \cite{XYModelFM5} discusses the use of Flow Matching as a scalable generative modeling framework for the XY Model.

Despite the success of SGMs, their standard formulation is inherently on the Euclidean embedding space. To formally adapt diffusion processes to the intrinsic geometry of data, \cite{RSGM2} introduced Riemannian Score-Based Generative Models (RSGM). This development proved all the more essential, as demonstrated by \cite{TangoDm}, whose analysis revealed that training on the Euclidean embedding space forces the model to prioritize learning the manifold constraint itself, compromising its ability to accurately learn the target distribution within the manifold.

In this work, we first provide extensive empirical evidence that training on the Euclidean embedding space leads to inaccurate estimates of thermodynamic quantities and the score function for the XY model. We address these limitations by proposing the application of a Manifold-Aware Score-Based Generative Modeling framework to the high-dimensional setting of a $64 \times 64$ 2D XY lattice (a 4096-dimensional torus). We also provide a white-box analysis by directly comparing the theoretical Boltzmann score with our model’s predictions, demonstrating superior precision compared to standard Euclidean diffusion models. This is particularly important as it provides a deeper insight into the training dynamics, moving beyond the black-box nature of such deep learning architectures. Such an interpretability analysis is of particular interest to researchers \cite{ScoreComparison13}. Since the XY model from statistical physics exhibits non-trivial topological features and phase transitions, it serves as a rigorous testbed for such an analyses. Notably, we accurately match second-moment quantities like heat capacity for the first time, while strictly adhering to finite-size scaling. By explicitly respecting manifold topology, this work provides a versatile methodology that extends beyond our primary focus on the XY model, offering a robust and interpretable approach for the generative modeling of continuous spin systems.

\section{Background and Related Literature}
The XY Model spin system in statistical physics allows spins to be oriented in any direction in the two-dimensional plane. In the XY Model, the Mermin-Wagner theorem rules out the existence of a long-range-ordered (LRO) phase in two dimensions \cite{merminWagner23}. However, the formation of topological defects (i.e., vortices and antivortices) results in a quasi-LRO phase, leading to a Berezinskii-Kosterlitz-Thouless (BKT) transition due to the vortex-antivortex unbinding above the critical temperature \cite{xymodelMAIN_22}. The system is governed by the following Hamiltonian:
\begin{equation}
    H = -J \sum_{\langle i, j \rangle} \cos(\theta_i - \theta_j)
\end{equation}
where $\theta_i, \theta_j \in [0, 2\pi)$ represents the spin angle on the lattice site $i$ and $j$ such that $i$ and $j$ are nearest-neighbours to each other, and $J$ is the ferromagnetic coupling constant (set to $J=1$ for our simulations). We consider only nearest-neighbor interactions. The system's probability distribution follows the Boltzmann distribution $p = \frac{1}{Z} e^{-\beta H}$, where $\beta = \frac{1}{k_B T}$, $Z$ is the partition function, $k_B$ is the Boltzmann constant, and $T$ is the temperature.
The helicity modulus (or spin stiffness) $\Upsilon$ helps us in identifying the phase transition. It is calculated from the fluctuations of the spin components as:
\begin{equation}
    \Upsilon = \frac{1}{N} \left( \langle h_y \rangle - \frac{1}{T} \langle I_y^2 \rangle \right)
\end{equation}
where $\langle h_y \rangle = \langle \sum_{\langle i, j \rangle_y} \cos(\theta_i - \theta_j) \rangle$ is the negative hamiltonian, and $I_y = \sum_{\langle i, j \rangle} \sin(\theta_i - \theta_j)$ represents the spin current as defined in \cite{sandvik25}. The critical temperature $T_{KT}$ is determined by the intersection of the measured $\Upsilon(T)$ curve with the straight line $\frac{2}{\pi} k_B T$.
The critical temperature $\tilde{T}_{KT}(L)$ observed in a finite system of size $L$ scales toward the thermodynamic limit $T_{KT}$ as given below: 
\begin{equation}
    \tilde{T}_{KT}(L) \approx T_{KT} + \frac{\pi^2}{4c(\ln L)^2}
\end{equation} where $c$ is a constant \cite{FSS24}. 

We also analyze the specific heat capacity, $C_v$, which is computed as: 
\begin{equation}
    C_v = \frac{\text{Var}(E)}{N T^2} = \frac{\langle E^2 \rangle - \langle E \rangle^2}{L^2 T^2}
\end{equation}

\subsection{Langevin Dynamics Simulation in XY Model}
Since the XY model is a continuous-spin system, it is well-suited for simulations with Langevin Dynamics, as detailed in \cite{LangevinSimulationXYModel12}. In their analysis of the model's dynamics, they derive equations for the overdamped limit (high viscosity), where the inertial term becomes negligible compared to the damping and force terms. This reduction leads to a dimensionless first-order Langevin equation. By adopting unit coupling ($J=1$) and Boltzmann constant ($k_B=1$), the equation governing the time evolution of the spin at site $i$ becomes:
\begin{equation}
    \frac{d\theta_i}{dt} = \sum_{\langle i, j \rangle} \sin(\theta_i - \theta_j) + \eta_i(t)
\end{equation}
Here, $\eta_i(t)$ represents the stochastic noise satisfying the fluctuation-dissipation theorem $\langle \eta(t)\eta(s) \rangle = 2T\delta(t-s)$.
In our implementation, we use the discretized version of this stochastic differential equation to generate proposals for the Metropolis-Adjusted Langevin Algorithm (MALA) \cite{MALA16} step in our sampling approach after replacing $\frac{1}{T}\sum_{\langle i, j \rangle} \sin(\theta_i - \theta_j)$ with the learned score function conditioned on temperature $T$. The standard MALA proposal with a time step $\tau$ for a target distribution $p(\theta) \propto e^{-\beta H}$ is given by $\theta_{\text{new}} = \theta + \tau \nabla \log p(\theta) + \sqrt{2\tau} z$, where $z \sim \mathcal{N}(0, I)$. For the XY model, we have $\nabla \log p(\theta) = \frac{1}{T} \sum_{\langle i, j \rangle} \sin(\theta_i - \theta_j)$. If we define a temperature-dependent step size $\tau = \epsilon T$, the standard proposal transforms into the discretized form:
\begin{equation}
    \theta_{\text{i,new}} = \theta_{\text{i,old}} + \epsilon \sum_{\langle i, j \rangle} \sin(\theta_{i,old} - \theta_j) + \sqrt{2 \epsilon T} z
\end{equation}

\subsection{Score-based Generative Modeling}
Score-based generative models (SGMs) circumvent the intractable normalizing constant of likelihood-based models by directly learning the score function, $\nabla_x \log p(x)$. For accurate estimation of the score function across the entire data manifold, Noise Conditional Score Network (NCSN) was introduced in \cite{SMLD6}. Training NCSN requires the data to be perturbed with a sequence of Gaussian noise scales $\sigma_1 < \sigma_2 < \dots < \sigma_N$. The approach trains a single conditional network (the NCSN model), $s_\theta(x, \sigma)$, to estimate the score of each perturbed distribution.
The training process uses the Denoising Score Matching (DSM) objective as given below:
$$\mathcal{L}(\theta) = \frac{1}{N}\sum_{i=1}^N \sigma_i^2 \mathbb{E}_{x \sim p_{data}, \tilde{x} \sim \mathcal{N}(x, \sigma_i^2 I)} \left[ \left\| s_\theta(\tilde{x}, \sigma_i) + \frac{\tilde{x} - x}{\sigma_i^2} \right\|^2 \right]$$
This objective encourages the network to recover the clean data $x$ from the noisy sample $\tilde{x}$.
This discrete multi-noise scale approach was subsequently unified into a continuous-time framework, which represented it as Score-based Stochastic Differential Equations (SDEs) \cite{ScoreSDE7}. In this unified view, the NCSN method discussed in \cite{SMLD6} corresponds to the Variance Exploding (VE) SDE. This contrasts with the Variance Preserving (VP) SDE, which describes Denoising Diffusion Probabilistic Models (DDPM) \cite{DDPM11}. While these continuous frameworks exist, our work uses the original discrete regime. However, we diverge from the standard annealed Langevin dynamics sampling. Instead, we employ an ODE sampling method to generate samples, followed by the Metropolis-Adjusted Langevin Algorithm (MALA). The MALA step is utilized to tune the variance of the generated distribution, thereby correcting the discretization errors inherent in the ODE sampling process.

\subsection{Generative Modeling in Spin Systems}
Generative modeling in spin systems aims to capture the underlying physics and statistical distribution of lattice configurations to generate physically accurate samples. Early approaches extensively explored deep learning techniques such as Restricted Boltzmann Machines (RBMs), Variational Auto-Encoders (VAEs), and Generative Adversarial Networks (GANs) \cite{SinghcGAN10, vae27,vae28,rbmIsing_26}. However, the rapid success of diffusion models in computer vision has motivated recent comparative studies on their thermodynamic fidelity relative to these predecessors. \cite{DiffusionBeatsGAN15} Diffusion models demonstrate superior capability in learning accurate data descriptions, primarily due to their mathematical robustness and stability, avoiding the training volatility often associated with adversarial objectives in GANs, specially for spin systems \cite{IsingDiffusion3}.

Building on this, recent work has employed Flow Matching \cite{XYModelFM5} to achieve improved generation quality, demonstrating strong zero-shot generalization and the ability to reproduce helicity modulus scaling across lattice dimensions. Despite these advancements, these methods typically train models within the Euclidean embedding space. To address inaccuracies in second-moment quantities like heat capacity, prior work \cite{thapar20} suggests that introducing specific regularization terms can yield significant improvements. In contrast, our work demonstrates that manifold-aware training is intrinsically sufficient to achieve high accuracy estimation of these second-moment quantities directly from the generated samples, without the need for an auxiliary regularization.
\subsection{Manifold Aware Training in Diffusion Models}
Standard diffusion models trained in Euclidean space face significant challenges when the underlying data support lies on a lower-dimensional manifold. In this setting, the score function $\nabla \log p_t(x)$ exhibits a multiscale singularity as the noise scale $\sigma$ approaches zero. Specifically, Ref. \cite{TangoDm} analyzes this by decomposing the score into tangential and normal components relative to the manifold, demonstrating that while the tangential component remains bounded ($O(1)$), the normal component diverges as $O(1/\sigma)$. Hence, for an accurate estimation of the score of the distribution within the manifold, it is important to perform the training and sampling on the manifold itself. An approach discussed in the literature for manifold-aware training of score-based generative models is the Riemannian Score-Based Generative Models (RSGM) \cite{RSGM2}, which avoids the Euclidean embedding space entirely by defining the forward diffusion and reverse denoising processes directly on the Riemannian manifold, though this requires knowledge of the manifold's intrinsic geometry. 

\section{Method}
\subsection{Background}
The fundamental objective of generative modeling is to approximate the underlying probability distribution, $p_{\text{data}}(x)$, of the given dataset. For physical systems, the normalization constant $Z$ of the Boltzmann distribution $p(x) = \frac{1}{Z} e^{-\beta E(x)}$ is computationally intractable to estimate from the limited dataset. Score-based Generative Modeling (SGM) circumvents this difficulty by learning the score function rather than directly estimating the density \cite{score_removes_z_29}. The score function is defined as the gradient of the log-probability density with respect to the input:
\begin{equation}
    s_\theta(x) \approx \nabla_x \log p(x)
\end{equation} 
This approach eliminates the need to compute $Z$, as the gradient of the constant $Z$ with respect to $x$ is zero. After training the model to learn the score function, the general idea is to stochastically generate samples using Langevin Dynamics iteratively, that evolves a random noise vector toward high-density regions of the data distribution:
\begin{equation}
    x_{t+1} \leftarrow x_t + \frac{\epsilon}{2} \nabla_x \log p(x_t) + \sqrt{\epsilon} z_t
\end{equation}
where $\epsilon$ is the step size and $z_t \sim \mathcal{N}(0, I)$.
However, this requires accurate estimation of score for all possible lattice configurations due to noise perturbations, even though they might be rarely present in the dataset. However, score estimation is inaccurate in low-density regions where training data is scarce. A detailed analysis of this issue is presented in \cite{SMLD6}.
To resolve these issues, a Noise Conditional Score Network (NCSN) framework is used, as proposed in \cite{SMLD6}, in which the data is perturbed with a sequence of noise levels $\{\sigma_i\}$, that fills the low-density regions, ensuring the model receives a significant training signal everywhere. NCSN is a single network conditioned on the noise level, $s_\theta(x, \sigma)$, to estimate the scores of these smoothed distributions. 
Hence, the forward process can be described as adding noise to the initial data until it reaches a prior distribution $\pi (x)$. During sampling, this process is reversed, i.e., noise is gradually removed step-by-step from a lattice sampled from the prior distribution to reach the data distribution. Although the original work \cite{SMLD6} proposed an Annealed Langevin Dynamics approach with annealing of the noise levels, Ordinary Differential Equation (ODE) sampling approaches were also proposed in \cite{ScoreSDE7}. 
The ODE Sampling approach is discussed in greater detail in the Sampling sub-section, as it is the primary sampling technique for our use case. Additionally, our work trains a conditional NCSN, i.e., the model is also conditioned on the temperature of the dataset along with the noise level (which is given as a time $t$ parameter, that indexes the noise level) $s_\theta (x, t, T)$ so as to train the temperature ($T$) dependency of lattice distributions to the model along with the noise dependency.
\paragraph{Manifold-Awareness for the XY Model} The XY model dataset is a toroidal dataset. The manifold for an $L \times L$ lattice is $\mathbb{T}^{L^2} = S^1 \times \dots \times S^1$, rather than Euclidean space $\mathbb{R}^N$. Hence, the use of Gaussian noise is geometrically ill-defined as it pushes the samples off the manifold. Thus,  a Wrapped Normal distribution is used for the noise perturbation process to ensure that the forward process respects the $2\pi$-periodicity of the angular variables. The Wrapped Normal Distribution is given as:
\begin{equation}
    p_{WN}(\theta; \mu, \sigma) = \frac{1}{\sigma \sqrt{2\pi}} \sum_{k=-\infty}^{\infty} \exp\left( -\frac{(\theta - \mu + 2\pi k)^2}{2\sigma^2} \right)
\end{equation}
\subsection{Architecture}
We use a U-Net \cite{unet17} architecture, inspired from \cite{IsingDiffusion3} with modifications to respect certain conditions of our physics dataset.
\paragraph{Circular Padding} Standard Convolutional Neural Networks (CNNs) utilize zero-padding, which creates artificial boundaries at the image edges. To strictly enforce the periodic boundary conditions followed by the XY model within the network, we use circular padding. This ensures that convolution operations wrap around the lattice edges.
\paragraph{Input Representation} Directly training the model on raw angular values $\theta \in [0, 2\pi)$ can potentially become inefficient as the model may not understand the periodic nature of angles. Hence, we embed the angles into a two-channel tensor by converting the angle at each lattice site to $(\sin \theta, \cos \theta)$. The network outputs a single channel representing the score.
We also maintain an Exponential Moving Average (EMA) of the model parameters with hyperparameter $\beta = 0.9999$.

\subsection{Training}
The model is trained to learn the score function of the perturbed data distribution on the toroidal manifold. The data preparation and augmentation strategy is inspired from \cite{XYModelFM5} paper. The dataset comprises 1500 lattice configurations ($64 \times 64$) sampled via the Metropolis algorithm across a temperature range of $0.1$ to $1.9$ at a temperature difference of $0.01$. Hence, the dataset comprises 1500 lattice configurations, each for 190 temperature values. We further oversampled the data for the ordered phase of the dataset, including the region of phase transition.
\paragraph{Data Augmentation} Data augmentation techniques aim to encourage the model to learn the underlying physics without overfitting the training data.  The energy of the XY model is determined by the cosine of the angle differences between neighbouring spins:
\begin{equation}
    H(\{\theta\}) = -J \sum_{\langle i, j \rangle} \cos(\theta_i - \theta_j) 
\end{equation}
This Hamiltonian allows the rotation of every spin in the lattice by a constant scalar $\phi$ leaving the energy (and thus the probability density) invariant.
During training, we apply a Random Global Rotation to each batch. For every sample $x$ in a batch, we sample a random phase shift $\phi \sim \text{Uniform}[0, 2\pi)$ and transform the input:
\begin{equation}
    x' = (x + \phi) \pmod{2\pi}
\end{equation}
This procedure significantly increases the diversity of the training data and explicitly encourages the model to learn that the score function is rotationally invariant.
\paragraph{Noise Schedule} We perturb the data using a geometric sequence of noise levels $\{\sigma_i\}_{i=1}^N$ with $N=147$. This corresponds to the forward noising process in score-based models \cite{SMLD6,ScoreSDE7,DDPM11}. The noise levels span from $\sigma_{\min} = 0.01$ to $\sigma_{\max} = 3.0$. This noise is sampled from a Wrapped Normal distribution to ensure that the lattice angles remain on the manifold. 
\paragraph{Loss Function} We utilize Denoising Score Matching (DSM) \cite{SMLD6} adapted for periodic coordinates as given below:
\begin{equation}
    \mathcal{L}(\theta) = \mathbb{E}_{t, x_0, \tilde{x}} \left[ \lambda(\sigma_t) \left\| s_\theta(\tilde{x}, \sigma_t) - \nabla_{\tilde{x}} \log p_{\text{wrapped}}(\tilde{x} | x_0) \right\|^2 \right]
\end{equation}
where $\lambda(\sigma) \propto \sigma^2$ for our case.
\subsection{Sampling}
Our sampling procedure employs a two-step approach. Initially, we generate samples using the deterministic Probability Flow ODE, and then refine them using a Metropolis-Adjusted Langevin Algorithm (MALA) to ensure the final configuration strictly adheres to the target physical distribution.
\paragraph{Step 1: Probability Flow ODE (PF-ODE)} The forward and reverse processes of such SGM models correspond to a Langevin Equation. For every Langevin Equation, there exists a Fokker-Planck Equation that serves as a Probability Flow Ordinary Differential Equation (PF-ODE). The evolution of probability through the PF-ODE is the same as the evolution of the probability distribution of the samples sampled using the reverse Stochastic Differential Equation (reverse-SDE). The details of this approach are discussed in \cite{ScoreSDE7}. Our framework of using discretized noise values for the forward process corresponds to the Variance Exploding (VE) SDE case discussed in this paper \cite{ScoreSDE7}. Hence, it allows us to use the PF-ODE sampling approach. 
For the Variance Exploding (VE) SDE, the PF-ODE is given by:
\begin{equation}
    d\mathbf{x}_t = -\frac{1}{2} \frac{d[\sigma^2(t)]}{dt} \nabla_\mathbf{x} \log p_t(\mathbf{x}) dt
\end{equation}
where $\sigma(t)$ represents the noise schedule and $\nabla_\mathbf{x} \log p_t(\mathbf{x})$ is the score function approximated by our neural network $s_\theta(\mathbf{x}, \sigma_t)$. ODE sampling requires fewer steps compared to the standard Annealed Langevin dynamics sampling technique. 
We employed the Third-order Runge-Kutta (RK3) method for ODE sampling. RK3 provides a higher-order approximation by evaluating the score function at three intermediate points within each time step to compute the final update, offering a superior trade-off between computational cost and trajectory accuracy.
\paragraph{Step 2: Metropolis-Adjusted Langevin Algorithm (MALA)} While the PF-ODE generates high-quality approximations, the discretization error inherent in numerical solvers can occasionally result in samples that deviate slightly from the true Boltzmann distribution in terms of second moment quantities like heat capacity. To correct this, we polish the generated samples using the Metropolis-Adjusted Langevin Algorithm (MALA). It is important to note that the manifold-aware score function learnt by our network allows us to use this approach. We explicitly use the learnt score function rather than the theoretical Hamiltonian, showing that our model has learnt the score function to high accuracy and is capable of generating physically correct samples all by itself.
Each MALA step consists of a proposal followed by an accept-reject decision:
\begin{enumerate}
    \item \textbf{Langevin Proposal:} We propose a new state $\tilde{\mathbf{x}}_{i+1}$ using the gradient information (score) from the trained model \cite{LangevinSimulationXYModel12}:
\begin{equation}
    \tilde{\mathbf{x}}_{i+1} = \mathbf{x}_i + \epsilon \nabla_\mathbf{x} \log p(\mathbf{x}_i) + \sqrt{2\epsilon} \mathbf{z}
\end{equation}
where $\mathbf{z} \sim \mathcal{WN}(0, I)$ is Wrapped Gaussian noise and $\epsilon$ is the step size with $\epsilon=\epsilon_0 T$. This ensures that the Langevin proposal step exactly matches the theoretically derived Langevin Equation for the XY Model in \cite{LangevinSimulationXYModel12}.
\item \textbf{Metropolis Correction:} To ensure the generated chain converges to the exact target distribution $p(\mathbf{x})$, the proposed update is accepted with probability $\alpha$:
\begin{equation}
    \alpha = \min \left( 1, \frac{p(\tilde{\bm{x}}_{i+1}) q(\bm{x}_i \mid \tilde{\bm{x}}_{i+1})}{p(\bm{x}_i) q(\tilde{\bm{x}}_{i+1} \mid \bm{x}_i)} \right)
\end{equation}
where $q(\mathbf{x}'|\mathbf{x})$ is the transition probability density of the Langevin proposal.
\end{enumerate}
\paragraph{Implementation Details} A critical challenge in score-based MALA is that our model provides the score $\nabla \log p(\mathbf{x})$ rather than the probability density $p(\mathbf{x})$ itself. Consequently, the energy difference $\Delta E = \log p(\tilde{\mathbf{x}}_{i+1}) - \log p(\mathbf{x}_i)$ required for the acceptance ratio is not directly available. We resolve this by numerically integrating the score function along the geodesic path on the torus connecting the current state $\mathbf{x}_i$ and the proposal $\tilde{\mathbf{x}}_{i+1}$.
Furthermore, to sample at arbitrary temperatures $T$ different from the training baseline (effectively $T=1.0$), we utilize a Scaled Score mechanism. The network's output is scaled by the inverse temperature ($1/T$) during both the proposal generation and integration steps, since the learnt score function at minimum noise essentially corresponds to a Boltzmann Distribution. This is done because the model’s accuracy is better at $T=1.0$ than at lower temperatures. We typically run this polishing procedure for $N=50$ steps.

\section{Results}

\subsection{Manifold-Aware vs. Manifold-Agnostic Diffusion}
This section provides a comprehensive comparison between the proposed Manifold-Aware Modeling and the standard Manifold-Agnostic Approach. We compare our approach with the DDPM model trained as proposed in \cite{IsingDiffusion3}. By analyzing score stability, thermodynamic observables, and topological configurations, we offer empirical evidence of the model's ability to learn the intrinsic physics of the XY model.

\subsubsection{Score Stability Analysis}
A critical advantage of the manifold-aware formulation is the stability of the score function during the reverse-diffusion process. In standard Euclidean diffusion models (DDPM), the data manifold $\mathcal{M}$ (here, the torus $\mathbb{T}^N$) is embedded in a higher-dimensional Euclidean space $\mathbb{R}^{2N}$. As the noise scale $\sigma \to 0$, the probability density concentrates sharply onto this lower-dimensional manifold \cite{TangoDm}. Consequently, the score function $\nabla \log p_t(\mathbf{x})$ develops a large component orthogonal to $\mathcal{M}$ to force the diffusion trajectories back onto the manifold.

\begin{figure}[htbp]
    \centering
    \begin{subfigure}[b]{0.48\textwidth}
        \centering
        \includegraphics[width=\textwidth]{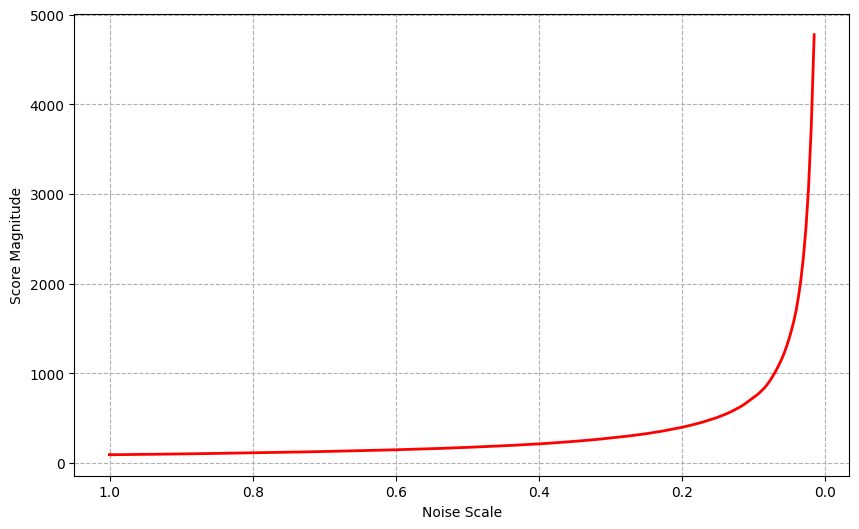}
        \caption{\textbf{Euclidean Space (DDPM)}}
        \label{fig:euclidean_score}
    \end{subfigure}
    \hfill
    \begin{subfigure}[b]{0.48\textwidth}
        \centering
        \includegraphics[width=\textwidth]{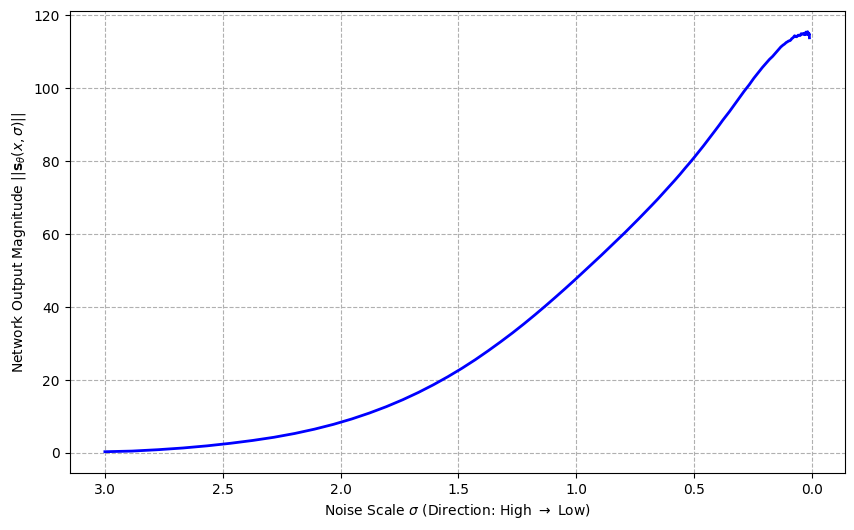}
        \caption{\textbf{Manifold-Aware}}
        \label{fig:manifold_score}
    \end{subfigure}
    \caption{\textbf{Comparison of Score Norm Scaling.} (a) In the standard Euclidean embedding, the score magnitude (red) explodes as $\mathcal{O}(1/\sigma)$ as $\sigma \to 0$. (b) In the proposed manifold-aware framework, the score magnitude (blue) remains $\mathcal{O}(1)$ and bounded.}
    \label{fig:score_comparison}
\end{figure}

The magnitude of this orthogonal component scales as $\|\mathbf{s}_\theta(\mathbf{x}, \sigma)\| \propto \mathcal{O}(1/\sigma)$, leading to numerical explosion at low noise levels. As shown in Figure \ref{fig:score_comparison}(a), the score magnitude for the model trained on Euclidean embedding space (DDPM model) diverges rapidly as $\sigma \to 0$. This forces the neural network to expend the majority of its capacity learning the geometry of the manifold (i.e., satisfying the constraint $\mathbf{x} \in \mathcal{M}$) rather than the distribution on the manifold.

In contrast, our manifold-aware approach defines the diffusion process intrinsically on the torus. The score vector lies entirely within the tangent space $T_{\mathbf{x}}\mathcal{M}$, eliminating the need for an orthogonal component. As illustrated in Figure \ref{fig:score_comparison}(b), the score magnitude in the manifold-aware model remains $\mathcal{O}(1)$ and bounded throughout the diffusion process. This stability ensures that the network focuses entirely on learning the intrinsic thermodynamics of the XY model, and not the manifold on which the data lies.

\subsubsection{Heat Capacity}
\begin{figure}[htbp]
    \centering
    \includegraphics[width=0.45\textwidth]{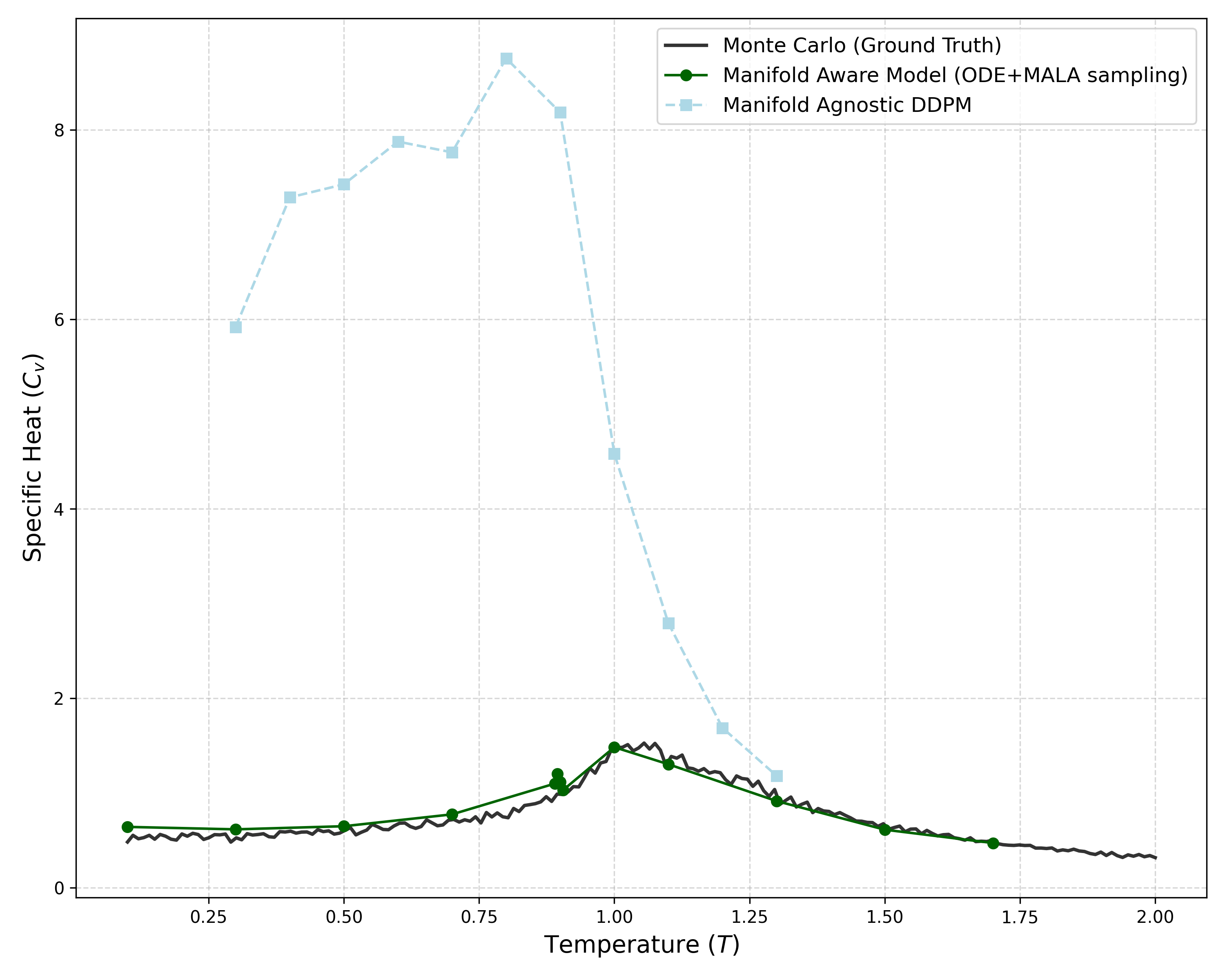} 
    \caption{\textbf{Specific Heat Capacity ($C_v$) vs. Temperature.} The Manifold-Aware model (green) reproduces the ground truth curve (black) and the heat capacity peak. The Manifold-Agnostic model (light blue) increases significantly at low temperatures.}
    \label{fig:specific_heat}
\end{figure}

We further evaluate the models by computing the specific heat capacity ($C_v$), a thermodynamic observable derived from energy fluctuations. Figure \ref{fig:specific_heat} compares the generated results against the Monte Carlo ground truth.

The manifold-aware model accurately tracks the ground truth across the entire temperature range. In contrast, the Manifold-Agnostic DDPM produces highly incorrect heat capacity values, particularly at lower temperatures ($T < 1.0$), where it fails to capture the correct energy variance.

\subsubsection{Spin Configurations and Topological Defects}
\begin{figure}[htbp]
    \centering
    \includegraphics[width=0.45\textwidth]{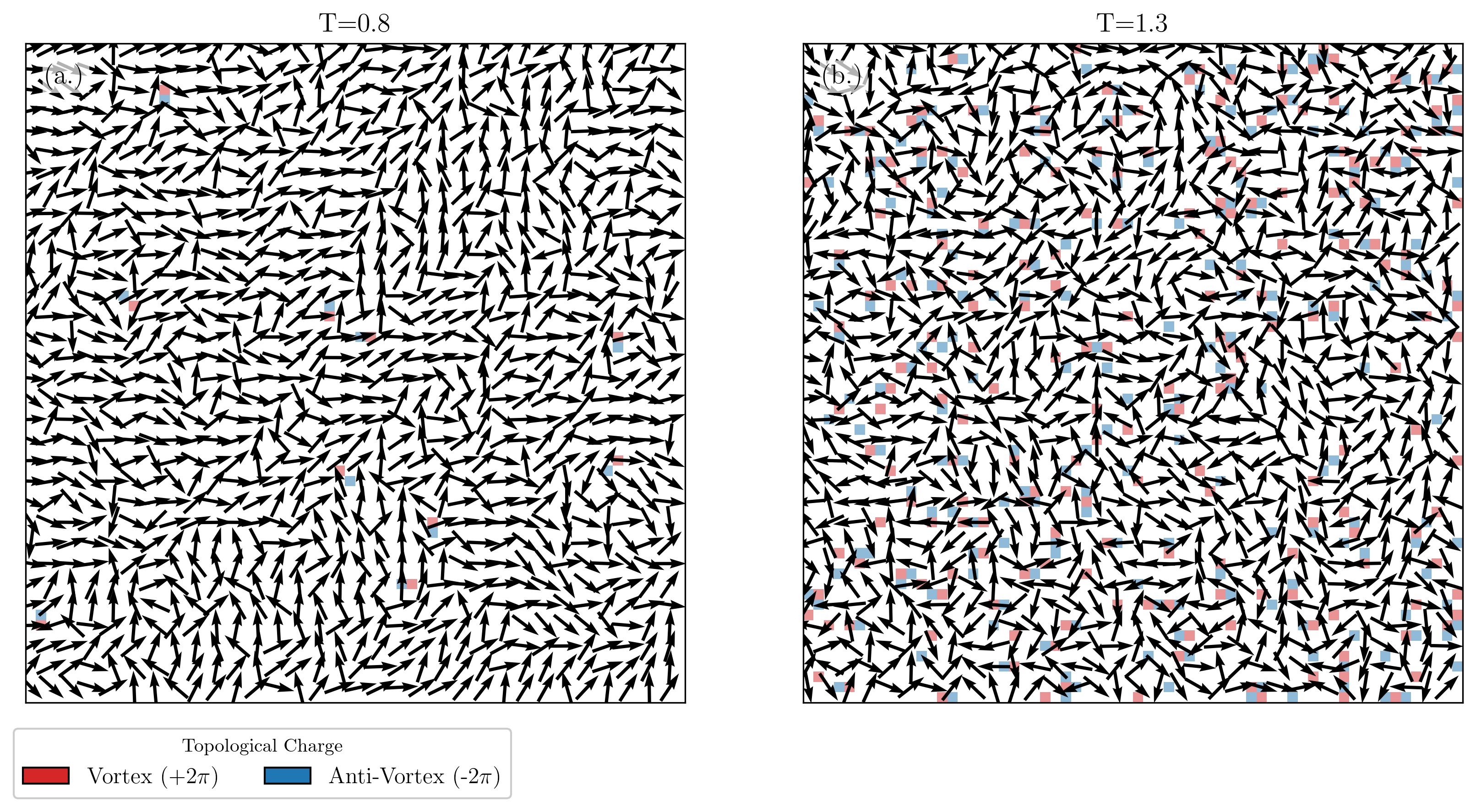}
    \caption{\textbf{Generated Spin Configurations.} (a) At $T=0.8$, the model generates bound vortex-antivortex pairs. (b) Above the critical temperature ($T=1.3$), the model correctly generates free vortices.}
    \label{fig:spin_configs}
\end{figure}

To verify that the model captures the global topology of the system, we examine the generated spin configurations. Figure \ref{fig:spin_configs} displays samples generated at temperatures below ($T=0.8$) and above ($T=1.3$) the critical point.

The visualization reveals that the model has successfully learned the underlying physics of vortex-antivortex unbinding, though not completely accurately. At $T=0.8$, the system exhibits bound vortex-antivortex pairs, characterizing the quasi-ordered phase. At $T=1.3$, these pairs unbind into free topological defects, characterizing the disordered phase. This behavior is further corroborated by the Helicity Modulus plot in Figure \ref{fig:helicity_modulus}.

\subsubsection{Helicity Modulus}
The Helicity Modulus ($\Upsilon$) serves as an order parameter for the XY model. Its behavior defines the critical temperature of the transition.

\begin{figure}[htbp]
    \centering
    \includegraphics[width=0.45\textwidth]{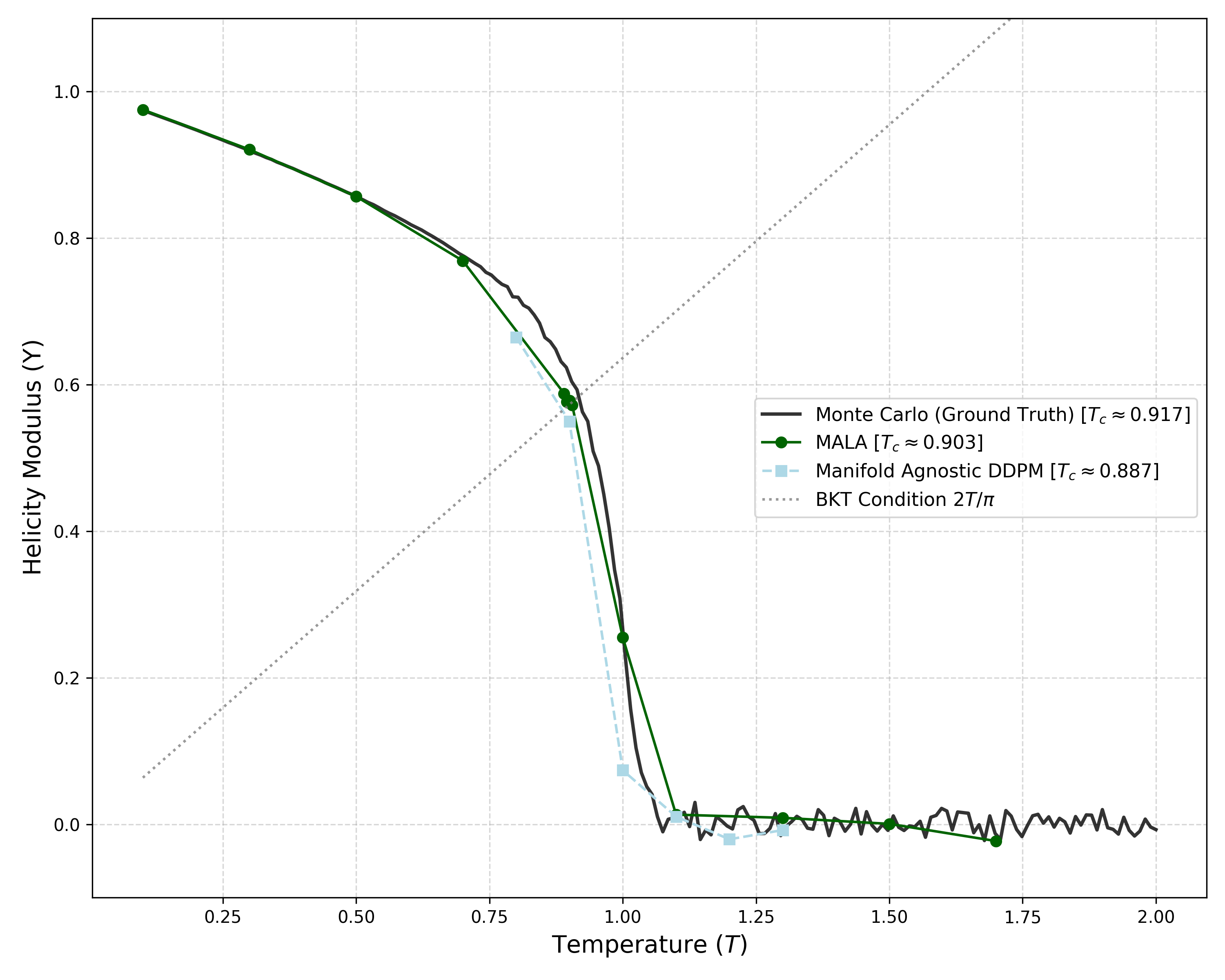}
    \caption{\textbf{Helicity Modulus ($\Upsilon$) vs. Temperature.} The Manifold-Aware model (green) accurately captures the stiffness jump at the critical temperature. The Manifold-Agnostic model (light blue) shows a lower critical temperature estimate.}
    \label{fig:helicity_modulus}
\end{figure}

Figure \ref{fig:helicity_modulus} illustrates the Helicity Modulus as a function of temperature. Though the Manifold-Aware model shows greater accuracy, the MALA results also closely follows the ground truth. This confirms that both the approaches exhibit tendency to correctly learn the global topological constraints of the system.

\subsubsection{Score Matching Verification}
Finally, we perform a white-box analysis by directly comparing the learned score function $\mathbf{s}_\theta(\mathbf{x})$ with the theoretical score derived from the XY model Hamiltonian.

\begin{figure}[htbp]
    \centering
    \includegraphics[width=0.95\textwidth]{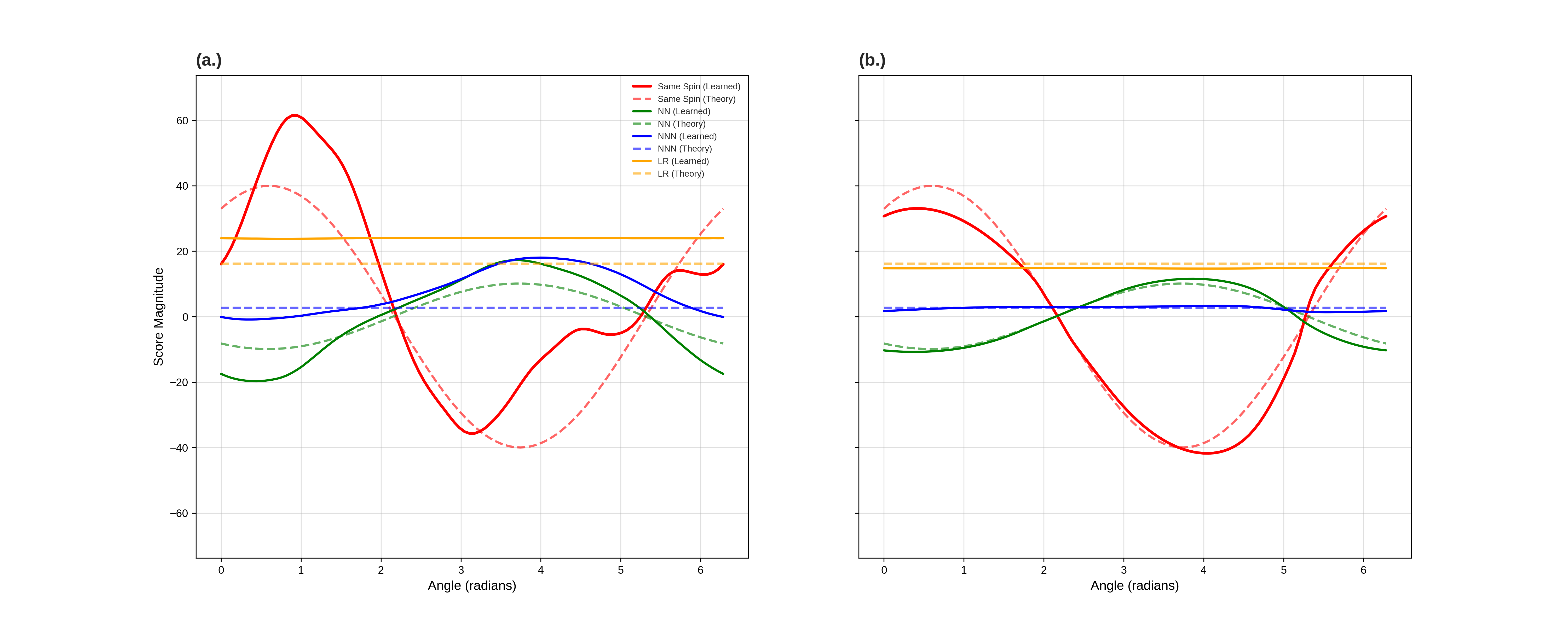}
    \caption{\textbf{Score Analysis.} Comparison of learned score components (solid) vs. theoretical values (dashed): (a.) Manifold-agnostic training of diffusion model, (b.) Manifold-aware training (our approach). This analysis has been done using a lattice at temperature$(T)=0.1$}
    \label{fig:score_matching}
\end{figure}

Figure \ref{fig:score_matching} presents the learned score magnitude versus the theoretical expectation for various interaction types. The substantial overlap between the learned curves (solid lines) and theoretical curves (dashed lines) for the manifold-aware model (Figure \ref{fig:score_matching}(a.)) confirms that the training objective has successfully aligned the network's output with the true vector field of the physical system. However, the manifold-agnostic DDPM model is unable to accurately capture the amplitude and form of the Hamiltonian (Figure \ref{fig:score_matching}(b.)). It also fails to correctly understand that the dataset exhibits only the nearest-neighbor interactions. This clearly shows that the manifold-agnostic approach is not able to correctly learn the distribution on the manifold.

\subsection{Hamiltonian Generalization and Temperature Scaling}

This section further serves as a direct visualization of whether the model has learnt the correct Hamiltonian, the temperature dependence, and the nearest-neighbour and long-range interactions of the XY Model. We select a fixed lattice configuration sampled from the low-temperature phase ($T_{\text{lattice}} = 0.1$) and manually perturb a central spin by rotating its angle $\theta$ from $0$ to $2\pi$. We then record the score magnitude predicted by the model at the minimum noise level for the central spin and its neighbours.

The theoretical score for the XY model is given by $\mathbf{s}(\mathbf{x}, T) = -\frac{1}{T} \nabla E(\mathbf{x})$. This implies that the score vector encodes two distinct pieces of information: the gradient of the hamiltonian ($\nabla E$) and a scalar coefficient ($1/T$).

\begin{figure}[!t]
    \centering
    \includegraphics[width=1.0\textwidth]{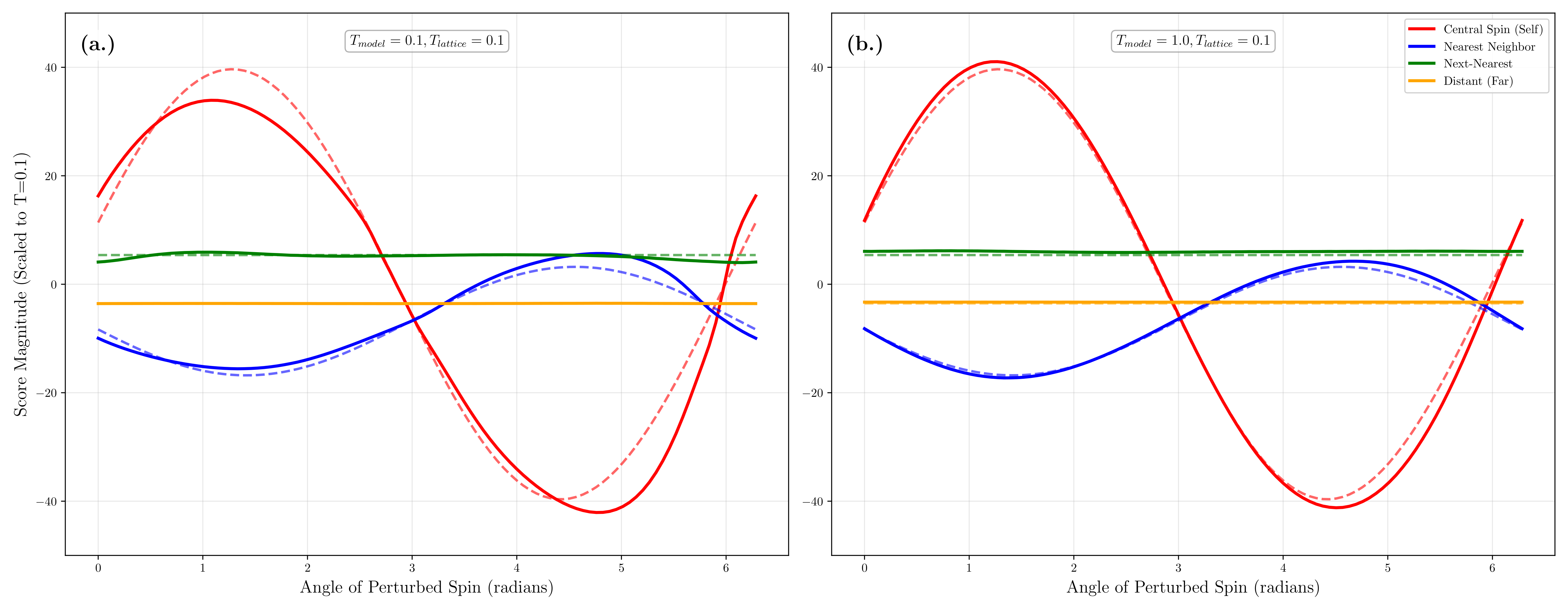}
    
    \caption{\textbf{Score estimation by temperature scaling.} 
    The score magnitude is plotted as a function of the angle of a single perturbed spin within a frozen $64 \times 64$ lattice ($T_{\text{lattice}}=0.1$). Solid lines represent model predictions, and the dashed lines represent the theoretical Hamiltonian forces.
    \textbf{(a.)} The model is conditioned on the matching temperature $T_{\text{model}}=0.1$.
    \textbf{(b.)} The model is conditioned on temperature $T_{\text{model}}=1.0$. The output scores are scaled by the ratio $T_{\text{model}}/T_{\text{lattice}}$.}
    
    \label{fig:cross_temp_probe}
\end{figure}

Figure \ref{fig:cross_temp_probe} presents the comparison between the model's predicted score (solid lines) and the analytical ground truth (dashed lines). The distinct lines correspond to different spatial interactions as given below:
\begin{itemize}
    \item \textbf{Red (Central Spin):} Represents the score function measured at the perturbed lattice site itself, responding to changes in its own angle. As expected from the Hamiltonian, this follows a sinusoidal form.
    \item \textbf{Blue (Nearest Neighbor):} Represents the score function measured for a nearest-neighbor lattice site due to the change in the central spin angle. This also correctly follows a sinusoidal graph.
    \item \textbf{Green (Next-Nearest) \& Orange (Distant):} These lines remain flat. This is a critical validation that the model has correctly learned that the interactions do not extend beyond immediate neighbors in the XY model, even though it was never explicitly constrained to be local.
\end{itemize}

In Figure \ref{fig:cross_temp_probe}(a), the model is conditioned on the correct temperature ($T_{\text{model}} = 0.1$). While it captures the general phase, the estimation shows slight variance.

In Figure \ref{fig:cross_temp_probe}(b), we test the model's generalization by conditioning it on a high temperature ($T_{\text{model}} = 1.0$) while inputting the same low-temperature lattice. Remarkably, when these high-temperature scores are scaled by the factor $T_{\text{model}} / T_{\text{lattice}}$, they yield a \textbf{better and more robust estimation} of the true score value than directly using $T=0.1$ Figure \ref{fig:cross_temp_probe}(a). This observation suggests that the model learns a smoother, globally consistent representation of the energy landscape $\nabla E$ at $T=1.0$. One potential reason for this can be because the model just above the critical temperature and hence, has more informed data at that temperature to correctly learn the score.

This result inspired us to use this scaling temperature strategy in our MALA polish step in sampling for the correct estimation of the acceptance ratio. This approach significantly improved the acceptance rates and stability of the MALA polish, as it provides a more accurate gradient estimate for the proposal distribution.

Though the score estimation by the model conditioned on $T=1.0$ gives more accurate results, it is important to understand the need to still train the conditional model conditioned for all temperature values. This is essential because we employ the score scaling strategy only for the MALA step. The first step of sampling the data using ODE sampler still requires the model to be trained on the required temperature value. This is because during ODE sampling, one traverses the whole high-noise to low-noise trajectory. The temperature scaling can only be done at the lowest noise level, since the model can be approximated to a boltzmann distribution only for this case. At other noise levels, the score learnt is of the noise-perturbed lattice, for which the temperature scaling is mathematically invalid.

\subsection{Vortex Structures in Sampled Lattices}

To verify that our generative model captures th non-trivial physics of vortex-antivortex unbinding at the BKT Transition, we visualized the raw spin configurations generated by the \textit{ODE + MALA} sampler from our manifold-aware model.

\begin{figure}[htbp]
    \centering
    \includegraphics[width=0.9\textwidth]{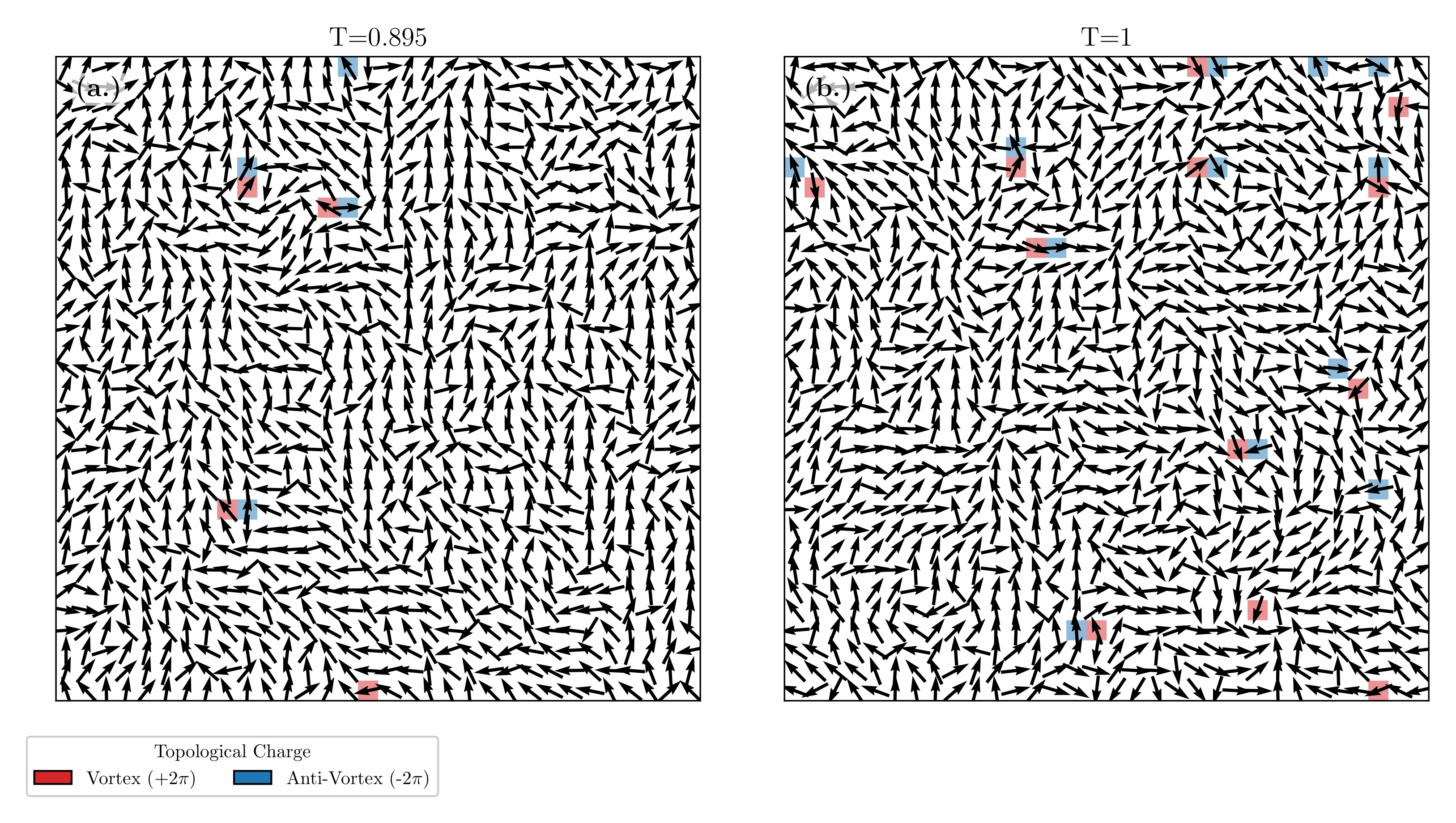}
    
    \caption{\textbf{Visualization of Sampled Spin Configurations.} 
    Lattice spins generated by the model at (a) $T=0.895$, slightly below the theoretical transition, and (b) $T=1.0$, slightly above the phase transition. The red squares correspond to vortices ($+2\pi$) and blue squares to anti-vortices ($-2\pi$). The clear formation of these defects demonstrates the model's ability to capture the BKT transition.}
    
    \label{fig:vortex_snapshots}
\end{figure}

Figure \ref{fig:vortex_snapshots} displays two lattices sampled from our model at temperatures $T=0.895$ (just below the critical point) and $T=1.0$ (in the disordered phase). We computed the local vorticity around each plaquette to identify vortices and anti-vortices. The vortex-antivortex pairs can be seen bound together at $T=0.895$, and the vortex-antivortex unbinding is also visible for the $T=1.0$ lattice.

\subsection{Helicity Modulus and the Universal Jump}

The universal jump in the helicity modulus ($\Upsilon$) is a characteristic of the XY Model.  $\Upsilon$ drops from a finite value to zero at the critical temperature $T_{BKT}$, satisfying the relation:
\begin{equation}
    \Upsilon(T_{BKT}) = \frac{2}{\pi} T_{BKT}
\end{equation}

\begin{figure}[htbp]
    \centering
    \includegraphics[width=1.0\textwidth]{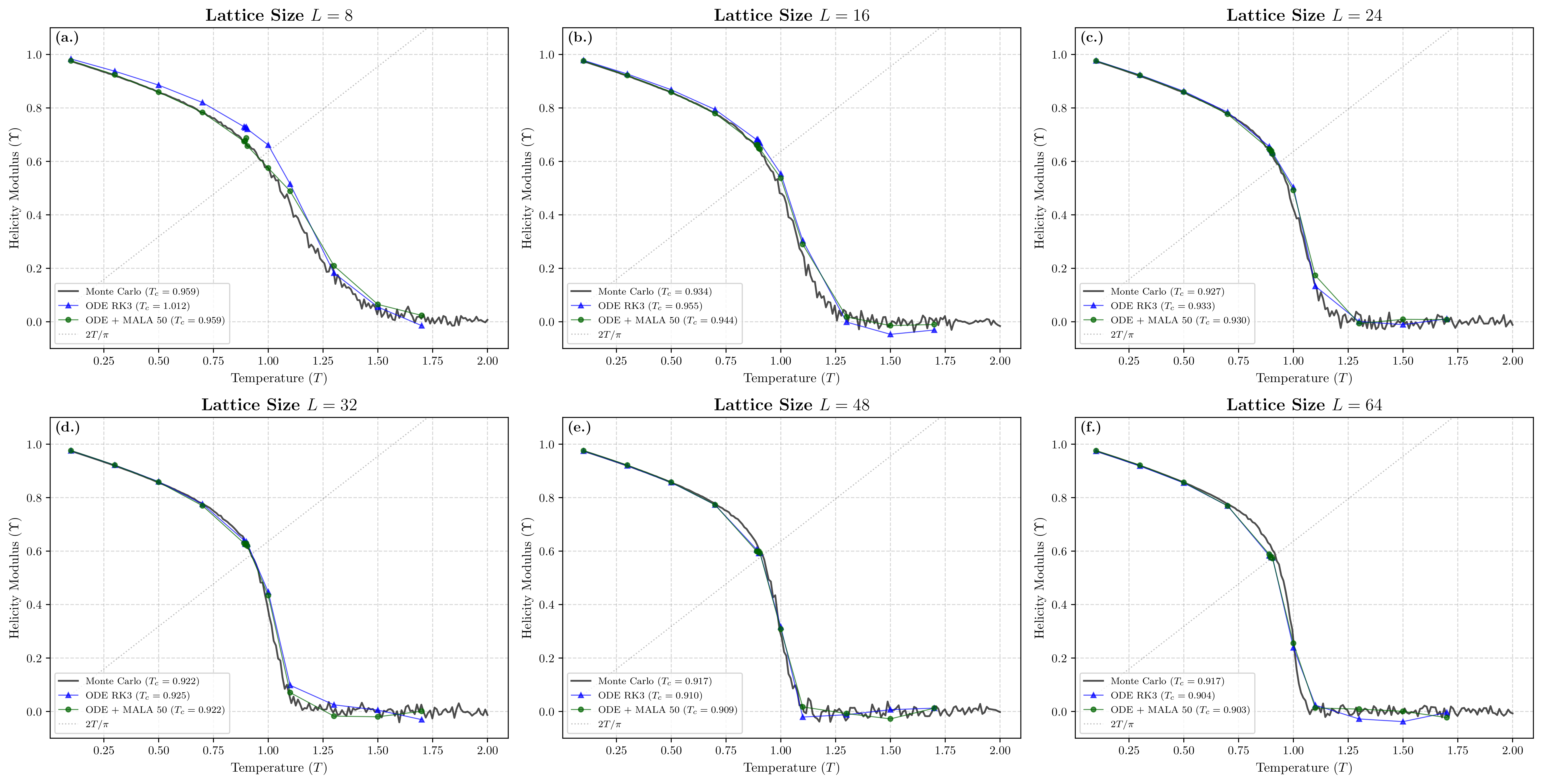}
    
    \caption{\textbf{Helicity Modulus $\Upsilon(T)$ Comparison.} 
    The helicity modulus is plotted as a function of temperature for lattice sizes $L \in \{8, 16, 24, 32, 48, 64\}$. The dashed diagonal line represents the universal jump condition $2T/\pi$. The intersection of the helicity modulus curves with this line denotes the critical temperature ($T_{BKT}$). The \textit{ODE + MALA} method shows a higher accuracy than only \textit{ODE}, specially for lower $L$ values. The model is trained only on the $L=64$ dataset. Hence, accurate prediction of helicity modulus for lower $L$ values depict good zero-shot generalization capabilities of our training approach.}
    \label{fig:helicity_modulus2}
\end{figure}

Figure \ref{fig:helicity_modulus2} displays the helicity modulus $\Upsilon(T)$ computed from our generated data from the model across different lattice sizes. The intersection of the measured $\Upsilon(T)$ with the straight line $2T/\pi$ provides an estimate for the transition temperature.  To test the model's zero-shot generalization capabilities, we used the trained score network to directly sample configurations for a range of unobserved lattice sizes $L \in \{8, 16, 24, 32, 48\}$ without any re-training or fine-tuning. As the lattice size increases from $L=8$ to $L=64$, the transition becomes sharper, and the intersection point converges towards the theoretical critical value. Notably, the \textit{ODE + MALA} samples show a greater agreement with the Monte Carlo ground truth, specially for lower values of $L$.

\subsection{Finite-Size Scaling}

To further rigorously assess the physical correctness of the generated samples, we performed a finite-size scaling analysis of the estimated the Berezinskii-Kosterlitz-Thouless (BKT) transition temperature. According to the theory for XY model, the critical temperature of the finite lattice $T_c(L)$ approaches the critical temperature for infinitely large lattice as:
\begin{equation}
    T_c(L) = T_{BKT} +  \frac{\pi^2}{4c(\ln L)^2}
\end{equation}

Our generative model was trained solely on lattice configurations of size $L=64$.

\begin{figure}[H] 
    \centering
    \includegraphics[width=0.7\textwidth]{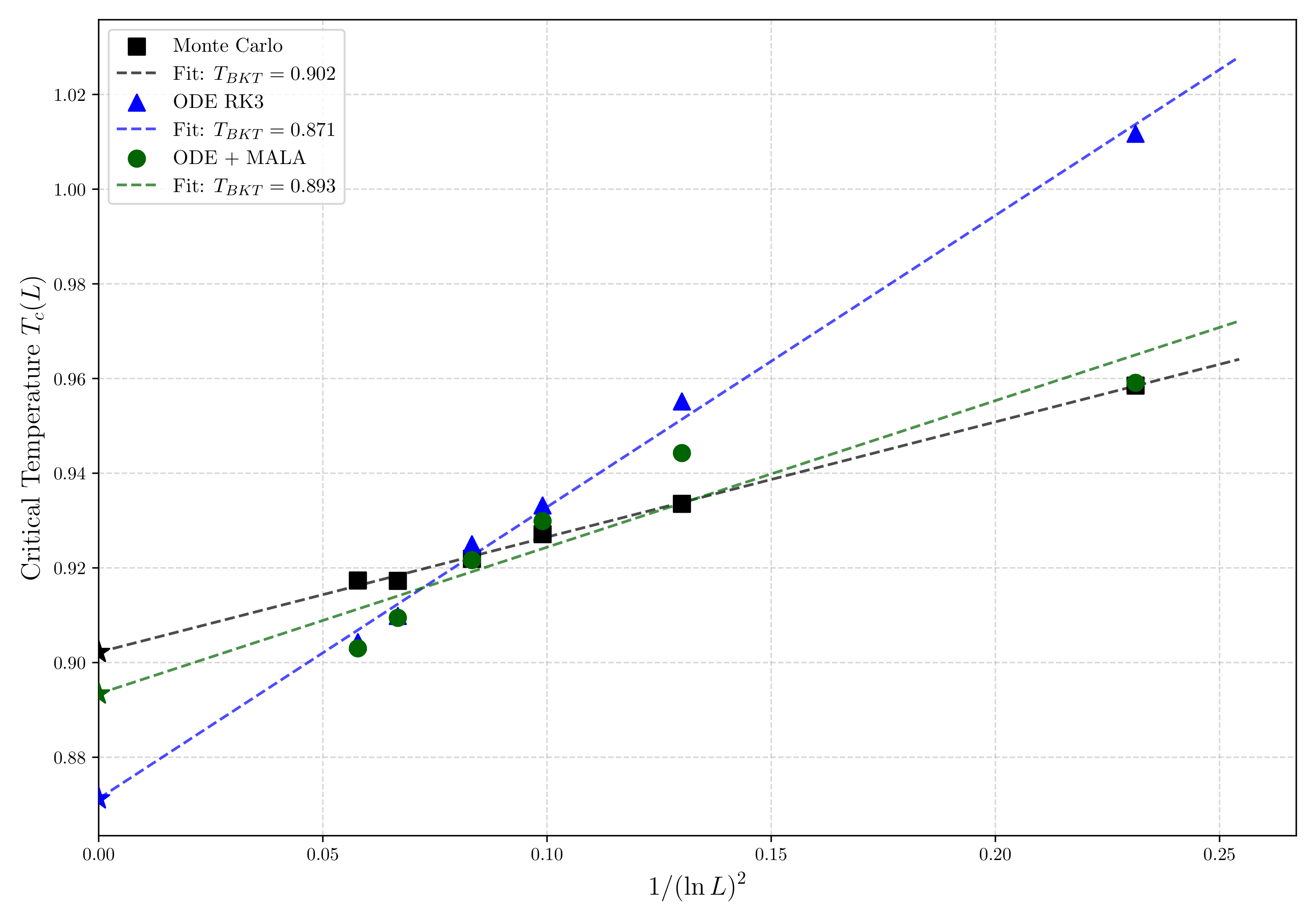}
    
    \vspace{-1em}
    
    \caption{\textbf{Finite-Size Scaling of the Critical Temperature.} 
    The critical temperature $T_c(L)$ plotted against $1/(\ln L)^2$ for lattice sizes $L \in \{8, 16, 24, 32, 48, 64\}$. The proposed \textit{ODE + MALA} method gives a closer estimation to that of the Monte Carlo reference.}
    
    \label{fig:fss_scaling}
\end{figure}

Figure \ref{fig:fss_scaling} illustrates the scaling of $T_c(L)$ against $1/(\ln L)^2$. Both sampling methods capture the expected linear scaling trend, confirming that the model has learned the underlying scale-invariant physics of the BKT transition. However, the \textit{ODE + MALA} technique demonstrates a higher accuracy as predicted. The extrapolation to the thermodynamic limit ($L \to \infty$) yields a critical temperature of $T_{BKT} \approx 0.893$ for the MALA-corrected samples, and $T_{BKT} \approx 0.871$ for the ODE sampler without MALA. The critical temperature from our Monte Carlo dataset is $T_{BKT} \approx 0.902$.  

These results further establishes the superiority of the proposed hybrid \textit{ODE + MALA} sampling strategy.

\subsection{Heat Capacity Comparison}

In this section, we evaluate the accuracy of the proposed generative model in estimating thermodynamic observables. Specifically, we compare the specific heat capacity $C_v$ derived from our sampling method (ODE + MALA) against the ground-truth Monte Carlo simulations and the baseline ODE RK3 solver.

\begin{figure}[htbp]
    \centering
    \includegraphics[width=1.0\textwidth]{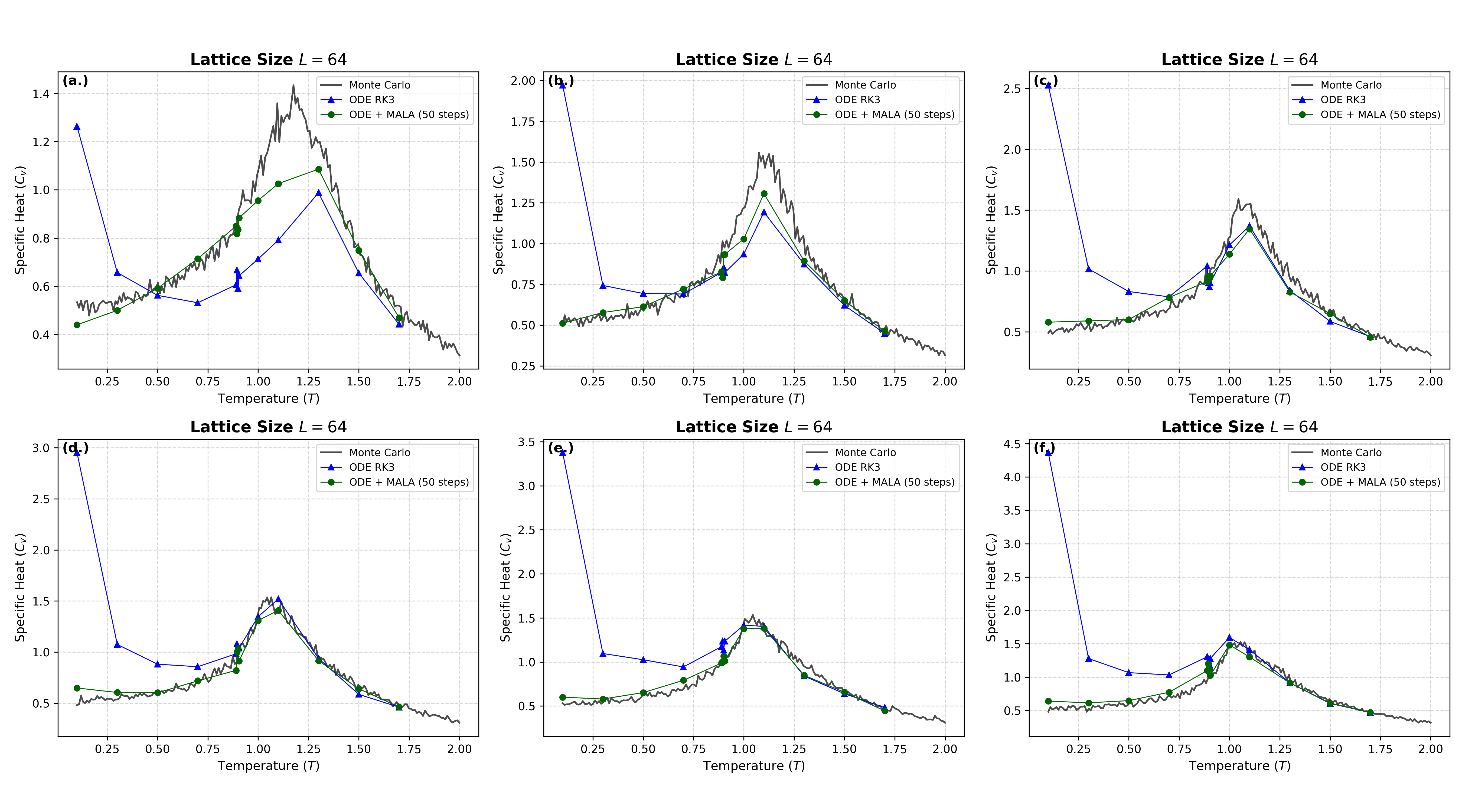}
    
    \caption{\textbf{Comparison of Specific Heat ($C_v$) across different lattice sizes.} 
    The plots display the specific heat capacity $C_v$ versus temperature $T$ for lattice sizes $L \in \{8, 16, 24, 32, 48, 64\}$. The black line represents the Monte Carlo reference data. The blue triangles denote the ODE RK3 method, and the green circles represent the proposed ODE + MALA (50 steps) approach.}
    \label{fig:cv_comparison}
\end{figure}

Figure \ref{fig:cv_comparison} presents the specific heat $C_v$ as a function of temperature $T$ across varying lattice sizes ($L$). The results demonstrate that the \textit{ODE + MALA} method shows better accuracy to the Monte Carlo baseline even at low temperatures ($T<1.0$). Without the MALA step, error accumulation due to discretization in ODE sampling leads to incorrect values, especially for lower temperatures. This result is significant, given that the heat capacity is a second-moment observable, and it shows that the correct sampling strategy is essential for accurately generating samples from the model.

As observed in Figure \ref{fig:cv_comparison}, the standard ODE RK3 method tends to deviate significantly at larger lattice sizes (e.g., $L=32$ and above) near the peak, whereas the hybrid MALA approach corrects these deviations effectively.

\subsection{Mean Energy per Site}
For the XY model, the mean energy per site, $\langle \epsilon \rangle$,  is calculated from the Hamiltonian as:
\begin{equation}
    \langle \epsilon \rangle = - \left\langle \frac{1}{L^2} \sum_{\langle i, j \rangle} \cos(\theta_i - \theta_j) \right\rangle
\end{equation}
where the sum runs over all nearest-neighbor pairs. As the temperature increases, the mean energy increases due to thermal disorder.

\begin{figure}[htbp]
    \centering
    \includegraphics[width=1.0\textwidth]{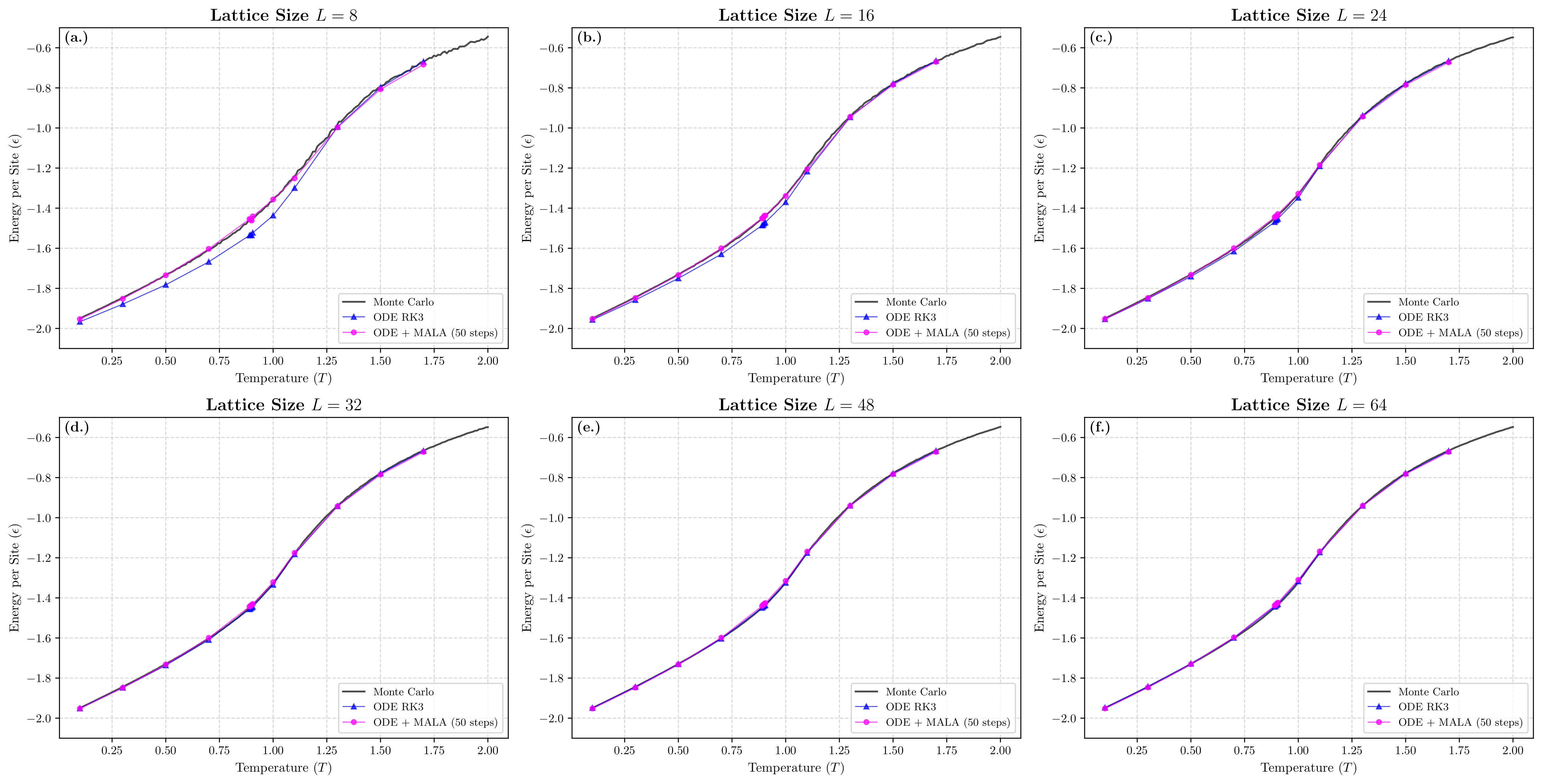}
    
    \caption{\textbf{Mean Energy per Site ($\langle \epsilon \rangle$) Comparison.} 
    The mean energy $\langle \epsilon \rangle$ is plotted as a function of temperature $T$ for lattice sizes $L \in \{8, 16, 24, 32, 48, 64\}$. The black line plot denotes the Monte Carlo ground truth.}
    
    \label{fig:mean_energy}
\end{figure}

Figure \ref{fig:mean_energy} compares the energy density of the generated samples against the Monte Carlo reference. The results show that the generative model faithfully reproduces the energy curve across all lattice sizes and temperatures. The $ODE+MALA$ sampling remains the more accurate approach, especially for lower values of lattice size.

\subsection{Magnetic Susceptibility}

\begin{figure}[htbp]
    \centering
    \includegraphics[width=1.0\textwidth]{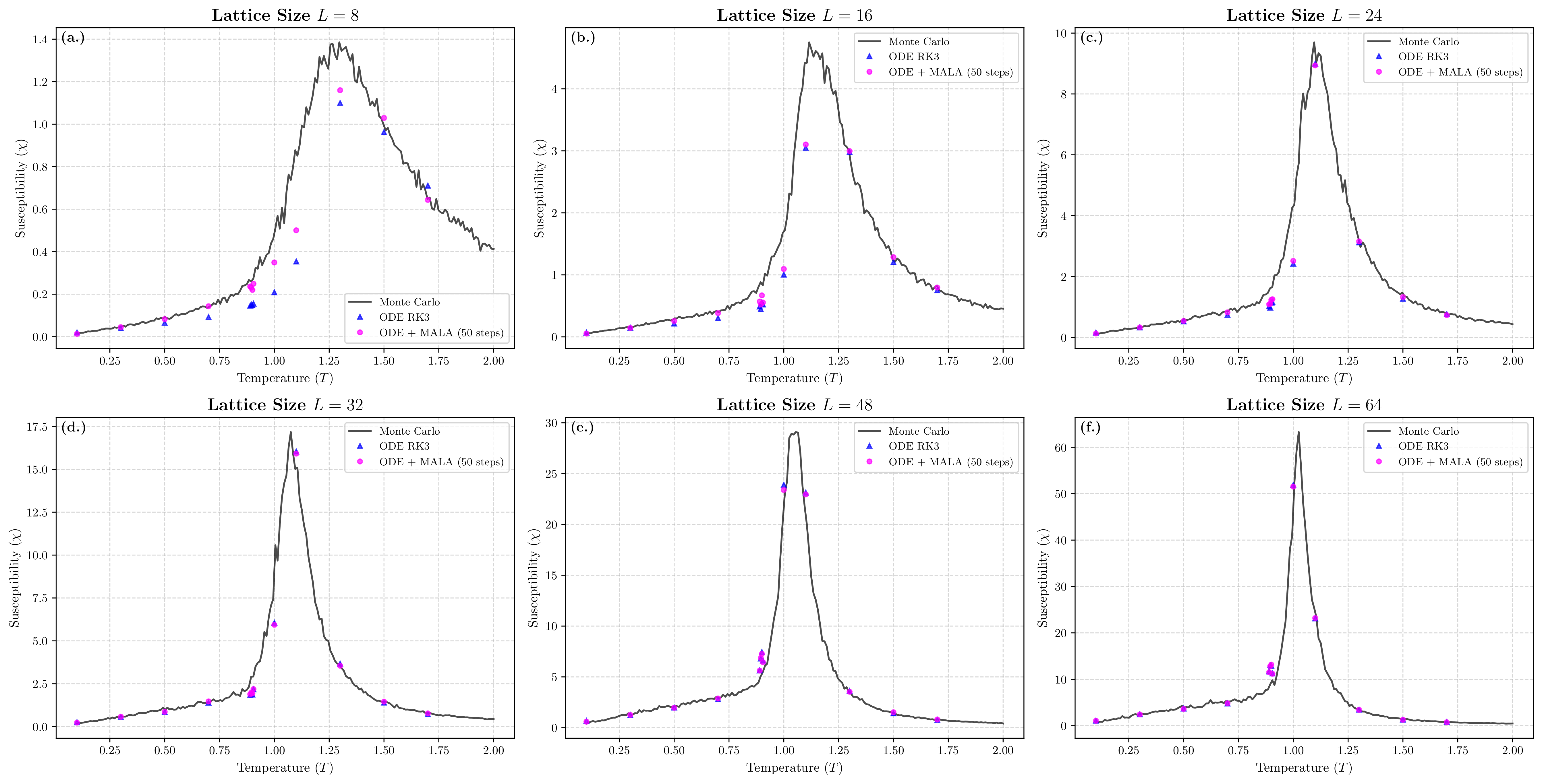}
    
    \caption{\textbf{Magnetic Susceptibility ($\chi$) Comparison.} 
    Susceptibility $\chi$ versus temperature $T$ for lattice sizes $L \in \{8, 16, 24, 32, 48, 64\}$. The black line plot denotes the Monte Carlo reference.}
    
    \label{fig:susceptibility}
\end{figure}

The susceptibility curves across varying lattice sizes is presented in figure \ref{fig:susceptibility}. The \textit{ODE + MALA} method demonstrates remarkable agreement with the Monte Carlo ground truth compared to the ODE solver (RK3), as predicted. The model is trained for $L=64$  is used to sample for all the other lattice size values.

\label{sec:headings}
\section{Conclusion}
In this work, we proposed a manifold-aware score-based generative framework for the 2D XY model. We demonstrated that respecting the intrinsic geometry effectively resolves the inconsistencies caused by training on the Euclidean embedding space. Our analysis confirms that the learned score aligns with the theoretical Boltzmann distribution, allowing us to accurately reproduce the BKT transition and the helicity modulus curve.

Crucially, we found that ODE sampling alone is insufficient for higher-order thermodynamics. We showed that subsequent MALA polishing is necessary to recover the specific heat capacity peak and correct discretization errors. Furthermore, the model's ability to generalize to unseen lattice sizes confirms that it has encoded the system's scale-invariant physics rather than simply memorizing the training data.

Looking forward, this framework serves as a blueprint for modeling more complex continuous symmetries. Natural extensions include applying this approach to analyze topological excitations like skyrmions, or to capture long-range ordering in complex many-body systems. Finally, our results encourage the exploration of similar frameworks for discrete systems (like the Ising model), solidifying the role of geometric deep learning as a rigorous tool for modern computational physics.

\section{Acknowledgement}
The author is grateful for discussions with Thakkar Meet Jiten.

\bibliographystyle{unsrt}
\bibliography{references}

\end{document}